\documentclass[10pt,twocolumn]{IEEEtran}
\usepackage[latin1]{inputenc}
\usepackage{amsmath}
\usepackage{amsbsy}
\usepackage{amssymb}
\usepackage{latexsym,color}
\usepackage{times}
\usepackage{epsfig}
\usepackage{amsfonts}
\usepackage{graphicx}
\usepackage{epstopdf}
\usepackage{cite}
\begin{document}
\title{Robust Adaptive Beamforming Based on Low-Complexity Shrinkage-Based Mismatch Estimation}
\author{Hang Ruan and Rodrigo C. de Lamare
\thanks{Copyright (c) 2012 IEEE. H. Ruan (Email: hang.ruan@york.ac.uk)
is with the Department of Electronics, University of York. R. C. de
Lamare (Email: delamare@cetuc.puc-rio.br) is with CETUC/PUC-Rio,
Brazil and with the Department of Electronics, University of York,
UK.}} \maketitle

\begin{abstract}
In this work, we propose a low-complexity robust adaptive
beamforming (RAB) technique which estimates the steering vector
using a Low-Complexity Shrinkage-Based Mismatch Estimation (LOCSME)
algorithm. The proposed LOCSME algorithm estimates the covariance
matrix of the input data and the interference-plus-noise covariance (INC)
matrix by using the Oracle Approximating Shrinkage (OAS) method.
LOCSME only requires prior knowledge of the angular sector in which
the actual steering vector is located and the antenna array
geometry. LOCSME does not require a costly optimization algorithm
and does not need to know extra information from the interferers,
which avoids direction finding for all interferers. Simulations show
that LOCSME outperforms previously reported RAB algorithms and has a
performance very close to the optimum.
\end{abstract}


\section{Introduction}

In applications like wireless communications, audio signal
processing, radar and microphone array processing, adaptive
beamforming has been intensively researched and developed in the
past years. However, under certain circumstances, adaptive
beamformers suffer a performance degradation due to several reasons
which include short data records, the presence of the desired signal
in the training data, or imprecise knowledge of the steering vector
of the desired signal. In order to improve the performance of
adaptive beamformers in the presence of steering vector mismatches,
RAB techniques have been developed. Different from the standard
designs \cite{r1}, the design principles of RAB MVDR beamformers
\cite{r6} include: the generalized sidelobe canceller, diagonal
loading \cite{r4,r5}, subspace projection
\cite{jio_el,jio_lcmv_esp,jidf_radar,jio_radar}, worst-case
optimization \cite{r3,r14} and steering vector estimation with
presumed prior knowledge \cite{r7,r8,r15,r16}. However, RAB designs
based on these principles have some drawbacks such as their ad hoc
nature, high probability of subspace swap at low SNR and high
computational cost \cite{r7}.

Some recent design approaches have considered combining different
design principles together to improve RAB performances. The
algorithm which jointly estimates the mismatched steering vector
using Sequential Quadratic Program (SQP) \cite{r8} and the
interference-plus-noise covariance (INC) matrix using a shrinkage
method \cite{r10} has been reported recently. Later, another similar
approach which jointly estimates the steering vector using SQP and
the INC matrix using a covariance reconstruction method \cite{r11},
presents outstanding performance compared to other RAB techniques.
However, the cost of the algorithm in \cite{r11} is high due to the
required matrix reconstruction process.

In this paper, we propose an RAB algorithm with low complexity,
which requires very little in terms of prior information, and has a
superior performance to previously reported RAB algorithms. The
proposed technique estimates the steering vector using a
Low-Complexity Shrinkage-Based Mismatch Estimation (LOCSME)
algorithm. LOCSME estimates the covariance matrix of the input data
and the INC matrix using the Oracle Approximating Shrinkage (OAS)
method. The only prior knowledge that LOCSME requires is the angular
sector in which the desired signal steering vector lies. Given the
sector, the subspace projection matrix of this sector can be
computed in very simple steps \cite{r7,r8,r9,r10,r11}. In the first
step, an extension of the OAS method \cite{r12} is employed to
perform shrinkage estimation for both the cross-correlation vector
between the received data and the beamformer output and the received
data covariance matrix. LOCSME is then used to estimate the
mismatched steering vector and does not involve any optimization
program, which results in a lower computational complexity. In a
further step, we estimate the desired signal power using the desired
signal steering vector and the received data. As the last step, a
strategy which subtracts the covariance matrix of the desired signal
from the data covariance matrix estimated by OAS is proposed to
obtain the INC matrix. The advantage of this approach is that it
circumvents the use of direction finding techniques for the
interferers, which are required to obtain the INC matrix.

This paper is structured as follows. The system model is described
in Section II. In Section III, the proposed LOCSME algorithm is
presented. Section IV shows and discusses the simulation results.
Finally, Section V gives the conclusion.

\section{System Model}

Consider a linear antenna array of $M$ sensors and $K$ narrowband
signals received at $i$th snapshot as expressed by
\begin{equation}
{\bf x}(i)={\bf A}({\boldsymbol \theta}){\bf s}(i)+{\bf n}(i),
\end{equation}
where ${\bf s}(i) \in {\mathbb C}^{K \times 1}$ presents the
uncorrelated source signals, ${\boldsymbol\theta}=[{\theta}_1,\dotsb,{\theta}_K]^T \in {\mathbb R}^K$ is the
vector containing the directions of arrivals (DoAs), ${\bf
A}({\boldsymbol \theta})=[{\bf a}({\theta}_1 )+{\bf e}, \dotsb, {\bf
a}({\theta}_K)] \in {\mathbb C}^{M \times K}$ is the matrix which
contains the steering vector for each DoA and ${\bf e}$ is the mismatch of
the steering vector of the desired signal, ${\bf n}(i) \in {\mathbb
C}^{M \times 1}$ is assumed to be complex Gaussian noise with zero
mean and variance ${\sigma}^2_n$. The beamformer output is given by
\begin{equation}
y(i)={\bf w}^H{\bf x}(i),
\end{equation}
where ${\bf w}=[w_1,\dotsb,w_M]^T \in {\mathbb C}^{M\times1}$ is the
beamformer weight vector, where $({\cdot})^H$ denotes the Hermitian
Transpose. The optimum beamformer can be computed by maximizing the
signal-to-interference-plus-noise ratio (SINR) given by
\begin{equation}
SINR=\frac{{\sigma}^2_1{\lvert{\bf w}^H{\bf a}\rvert}^2}{{\bf
w}^H{\bf R}_{i+n}{\bf w}}.
\end{equation}
Assume that the steering vector ${\bf a}$ is known precisely (${\bf
a}={\bf a}({\theta}_1 )$), where ${\sigma}^2_1$ is the desired
signal power and ${\bf R}_{i+n}$ is the INC matrix, then problem (3)
can be transformed into an optimization problem
\begin{equation}
\begin{aligned}
& \underset{\bf w} {\text{minimize}}
&& {\bf w}^H{\bf R}_{i+n}{\bf w} \\
& \text{subject to} && {\bf w}^H{\bf a}=1,
\end{aligned}
\end{equation}
which is known as the MVDR beamformer or Capon beamformer \cite{r1}.
The optimum weight vector is given by ${\bf w}_{opt}=
\frac{{{\bf R}^{-1}_{i+n}}{\bf a}}{{\bf a}^H{{\bf R}^{-1}_{i+n}}{\bf a}}.$
Since ${\bf R}_{i+n}$ is usually unknown in practice, it is estimated by the sample covariance matrix (SCM) of the received data as
\begin{equation}
\hat{\bf R}(i)=\frac{1}{i}\sum\limits_{k=1}^i{\bf x}(k){{\bf
x}^H}(k),
\end{equation}
which will result in the Sample Matrix Inversion (SMI) beamformer ${\bf
w}_{SMI}=\frac{\hat{\bf R}^{-1}{\bf a}}{{\bf a}^H\hat{\bf
R}^{-1}{\bf a}}$. However, the SMI beamformer requires a large
number of snapshots and is sensitive to steering vector mismatches
\cite{r10,r11}.

\section{Proposed LOCSME Algorithm}

In this section, the proposed LOCSME algorithm is introduced. The
idea of LOCSME is to estimate the steering vector and the INC matrix
separately as in previous approaches. The estimation of the steering
vector is described as the projection onto a predefined subspace
matrix of an iteratively shrinkage-estimated cross-correlation
vector between the beamformer output and the array observation. The
INC matrix is obtained by subtracting the desired signal covariance
matrix from the data covariance matrix estimated by the OAS method.

\subsection{{Steering Vector Estimation using LOCSME}}

The cross-correlation between the array observation vector and the beamformer output can be expressed as
\begin{equation}
{\bf d}=E\lbrace{\bf x}y^*\rbrace.
\end{equation}
We assume that ${\lvert{{\bf a}_m^H{\bf
w}}\rvert}\ll{\lvert{{\bf a}_1^H{\bf w}}\rvert}$ for $m=2,\dotsb,K$,
all signal sources and the noise have zero mean, and the desired
signal and every interferer are independent from each other. By
substituting (1) and (2) into (6), we suppose the interferers are
sufficiently canceled such that they fall much below the noise floor
and the desired signal power is not affected by the interference so
that ${\bf d}$ can be rewritten as
\begin{equation}
{\bf d}=E\lbrace{{{{\sigma}_1}^2{\bf a}_1^H{\bf w}{\bf a}_1}+{\bf n}{\bf n}^H{\bf w}}\rbrace.
\end{equation}
In order to eliminate the unwanted part of ${\bf d}$ and obtain an
estimate of the steering vector ${\bf a}_1$, ${\bf d}$ can be
projected onto a subspace \cite{r9} that collects information about
the desired signal. Here the prior knowledge amounts to providing an
angular sector range in which the desired signal is located, say
$[{\theta}_1-{\theta}_e,{\theta}_1+{\theta}_e]$. The subspace
projection matrix ${\bf P}$ is given by
\begin{equation}
{\bf P}=[{\bf c}_1,{\bf c}_2,\dotsb,{\bf c}_p][{\bf c}_1,{\bf
c}_2,\dotsb,{\bf c}_p]^H,
\end{equation}
where ${\bf c}_1,\dotsb,{\bf c}_p$ are the $p$ principal
eigenvectors vectors of the matrix ${\bf C}$, which is defined by
\cite{r8}
\begin{equation}
{\bf C}= \int
\limits_{{\theta}_1-{\theta}_e}^{{\theta}_1+{\theta}_e}{\bf
a}({\theta}){\bf a}^H({\theta})d{\theta}.
\end{equation}
At this point, LOCSME will use the OAS method to compute the
correlation vector ${\bf d}$ iteratively. The aim is to devise a
method that estimates ${\bf d}$ more accurately with the help of the
shrinkage technique. An accurate estimate of ${\bf d}$ can help to
obtain a better estimate of the steering vector. Let us define
\begin{equation}
\hat{\bf F}=\hat{\nu} {\bf I},
\end{equation}
where $\hat{\nu} = {{\rm tr}(\hat{\bf S})}/M$ and $\hat{\bf S} =
{\rm diag}({\bf x}y^*)$. Then, a reasonable tradeoff between
covariance reduction and bias increase can be achieved by shrinkage
of $\hat{\bf S}$ towards $\hat{\bf F}$ \cite{r12} and subsequently
using it in a vector shrinkage form, which results in
\begin{equation}
\hat{\bf d} = \hat{\rho}{\rm diag}(\hat{\bf F}) + (1-\hat{\rho}){\rm
diag}(\hat{\bf S}),
\end{equation}
which is parameterized by the shrinkage coefficient $\hat{\rho}$. If
we define $\hat{\bf D}={\rm diag}(\hat{\bf d})$ then the goal is to
find the optimal value of $\hat{\rho}$ that minimizes the mean
square error (MSE) of $E[{\lVert{\hat{\bf D}(i)-\hat{\bf
F}(i-1)}\rVert}^2]$ in the $i$th snapshot, which leads to
\begin{equation}
\hat{\bf d}(i) = \hat{\rho}(i){\rm diag}(\hat{\bf F}(i)) +
(1-\hat{\rho}(i)){\rm diag}(\hat{\bf S}(i)),
\end{equation}
\begin{equation}
\hat{\rho}(i+1) = \frac{(1-\frac{2}{M}){\rm tr}(\hat{\bf
D}(i)\hat{\bf S}^*(i)) + {\rm tr}(\hat{\bf D}(i)){\rm tr}(\hat{\bf
D}^*(i))}{(i+1-\frac{2}{M}){\rm tr}(\hat{\bf D}(i)\hat{\bf
S}^*(i))+(1-\frac{i}{M}){\rm tr}(\hat{\bf D}(i)){\rm tr}(\hat{\bf
D}^*(i))},
\end{equation}
where the derivation is shown in the Appendix and $\hat{\bf S}(i)$
is the sample correlation vector (SCV) given by
\begin{equation}
\hat{\bf S}(i)={\rm diag}\Big(\frac{1}{i}\sum\limits_{k=1}^i{\bf
x}(k)y^*(k)\Big).
\end{equation}
As long as the initial value of $\hat{\rho}(0)$ is between $0$ and
$1$, the iterative process in (12) and (13) is guaranteed to
converge \cite{r12}. Once the correlation vector $\hat{\bf d}$ is
obtained by the above OAS method, the steering vector is estimated by
\begin{equation}
{\hat{\bf a}_1}(i)=\frac{{\bf P}\hat{\bf d}(i)}{{\lVert{{\bf
P}\hat{\bf d}(i)}\rVert}_2},
\end{equation}
where $\hat{\bf a}_1(i)$ gives the final estimate of the steering
vector.

\subsection{{Interference-Plus-Noise Covariance Matrix Estimation}}

In order to estimate the INC matrix, the data covariance matrix
(which contains the desired signal) is required. The SCM in (5) is
necessary as a preliminary approximation. In the next step, similar
to using OAS to estimate the cross-correlation vector $\hat{\bf d}$,
the SCM is also processed with the OAS method as a further shrinkage
estimation step. Let us define the following quantity
\begin{equation}
\hat{\bf F}_0={\hat{\nu}_0}{\bf I},
\end{equation}
where $\hat{\nu}_0={{\rm tr}(\hat{\bf R})}/M$. Then, we use the
shrinkage form again
\begin{equation}
\tilde{\bf R}={\hat{\rho}_0}{\hat{\bf F}_0}+(1-{\hat{\rho}_0})\hat{\bf R}.
\end{equation}
By minimizing the MSE described by $E[{\lVert{{\tilde{\bf
R}(i)}-{\hat{\bf F}_0}(i-1)}\rVert}^2]$, we obtain the following recursion
\begin{equation}
\tilde{\bf R}(i)={\hat{\rho}_0}(i){\hat{\bf
F}_0}(i)+(1-{\hat{\rho}_0}(i))\hat{\bf R}(i),
\end{equation}
\begin{equation}
{\hat{\rho}_0}(i+1)=\frac{(1-{\frac{2}{M}}){\rm tr}({\tilde{\bf
R}(i)}{\hat{\bf R}(i)})+{\rm tr}^2(\tilde{\bf
R}(i))}{(i+1-{\frac{2}{M}}){\rm tr}({\tilde{\bf R}(i)}{\hat{\bf
R}(i)})+(1-{\frac{i}{M}}){\rm tr}^2(\tilde{\bf R}(i))}.
\end{equation}
Provided that $0<{\hat{\rho}_0}(0)<1$, the iterative process in (18)
and (19) is guaranteed to converge \cite{r12}. In order to eliminate
the unwanted information of the desired signal in the covariance
matrix and obtain the INC matrix, the desired signal power
${\sigma}^2_1$ must be estimated. Let us rewrite the received data
as
\begin{equation}
{\bf x}=\sum\limits_{k=1}^K{\bf a}_k{s_k}+{\bf n}.
\end{equation}
Pre-multiplying the above equation by ${\bf a}_1^H$, we have
\begin{equation}
{\bf a}_1^H{\bf x}={\bf a}_1^H{\bf a}_1{s_1}+{\bf a}_1^H(\sum\limits_{k=2}^K{\bf a}_k{s_k}+{\bf n}).
\end{equation}
Assuming that ${\bf a}_1$ is uncorrelated with the interferers, we
obtain
\begin{equation}
{\bf a}_1^H{\bf x}={\bf a}_1^H{\bf a}_1{s_1}+{\bf a}_1^H{\bf n}.
\end{equation}
Taking the expectation of $E[|{\bf a}_1^H{\bf x}|^2]$, we obtain
\begin{equation}
|{\bf a}_1^H{\bf x}|^2=E[({\bf a}_1^H{\bf a}_1{s_1}+{\bf a}_1^H{\bf n})^*({\bf a}_1^H{\bf a}_1{s_1}+{\bf a}_1^H{\bf n})].
\end{equation}
If the noise is statistically independent of the desired signal, we
have
\begin{equation}
|{\bf a}_1^H{\bf x}|^2=|{\bf a}_1^H{\bf a}_1|^2|s_1|^2+{\bf a}_1^H{\bf n}{\bf n}^H{\bf a}_1,
\end{equation}
where $|s_1|^2$ is the desired signal power which can be replaced by
its estimate $\hat{\sigma}^2_1$, ${\bf n}{\bf n}^H$ represents the
noise covariance matrix ${\bf R}_n$ which can be replaced by
${\sigma}^2_n{\bf I}_M$. Replacing ${\bf a}_1$ by its estimate
${\hat{\bf a}_1}(i)$ the desired signal power estimate is given by
\begin{equation}
\hat{\sigma}^2_1(i)=\frac{|{{\hat{\bf a}_1}^H}(i){\bf x}(i)|^2-{{\hat{\bf a}_1}^H}(i){\hat{\bf a}_1}(i){\sigma}^2_n}{|{{\hat{\bf a}_1}^H}(i){\hat{\bf a}_1}(i)|^2}.
\end{equation}
As the last step, the desired signal covariance matrix is subtracted and the INC matrix is given by
\begin{equation}
{\tilde{\bf R}_{i+n}}(i)=\tilde{\bf R}(i)-\hat{\sigma}^2_1(i){\hat{\bf
a}_1}(i){{\hat{\bf a}_1}^H}(i).
\end{equation}
The advantage of this step compared to SMI and existing methods is
that it does not require direction finding and is suitable for
real-time applications. With the estimates for the steering vector
and the INC matrix, the beamformer is computed by
\begin{equation}
\hat{\bf w}(i)=\frac{{\tilde{\bf R}^{-1}_{i+n}}(i)\hat{\bf a}_1(i)}{{\hat{\bf a}_1}^H(i){\tilde{\bf R}^{-1}_{i+n}}(i)\hat{\bf a}_1(i)}.
\end{equation}
Table I summarizes LOCSME in steps. From a complexity point of view,
the main computational cost is due to the following steps: SCM of
the observation data, OAS estimation for SCM, norm computations of
the covariance matrix and the INC matrix. Each of these steps has a
complexity of ${\mathcal{O}}(M^3)$. Additionally, compared to the
previous RAB algorithms in \cite{r7}, \cite{r8}, \cite{r10} and \cite{r11} which
have complexity equal or higher than ${\mathcal{O}}(M^{3.5})$,
LOCSME has a lower cost (${\mathcal{O}}(M^3)$).

\begin{table}
\small
\begin{center}
\caption{Proposed LOCSME Algorithm} \vspace{-0.5em}
\begin{tabular}{|l|}
\hline
Initialize: \\
${\bf C}=\int\limits_{{\theta}_1-{\theta}_e}^{{\theta}_1+{\theta}_e}{\bf a}({\theta}){\bf a}^H({\theta})d{\theta}$ \\
$[{\bf c}_1,\dotsb,{\bf c}_p]$: p princical eigenvectors of ${\bf C}$ \\
Subspace projection ${\bf P}=[{\bf c}_1,\dotsb,{\bf c}_p][{\bf c}_1,\dotsb,{\bf c}_p]^H$ \\
$\hat{\bf R}(0)={\bf 0}$; $\hat{\bf S}(0)={\bf 0}$; ${\bf w}(0)={\bf 1}$; \\
$\hat{\rho}(1)=\rho(0)=\hat{\rho}_0(1)=\rho_0(0)=1$; \\
For each snapshot index $i=1,2,\dotsb$: \\
$\hat{\bf R}(i)=\frac{1}{i}\sum\limits_{k=1}^i{\bf x}(k){{\bf x}^H}(k)$ \\
$\hat{\bf S}(i)= {\rm
diag}(\frac{1}{i}\sum\limits_{k=1}^i{\bf x}(k)y^*(k))$ \\
$\hat{\nu}(i) = {{\rm tr}(\hat{\bf S}(i))}/M$ \\
$\hat{\bf F}(i)=\hat{\nu}(i){\bf I}$ \\
$\hat{\bf d}(i) = \hat{\rho}(i) {\rm diag}(\hat{\bf F}(i)) +
(1-\hat{\rho}(i)){\rm
diag}(\hat{\bf S}(i))$ \\
$\hat{\bf D}(i)= {\rm
diag}(\hat{\bf d}(i))$ \\
$\hat{\rho}(i+1) = \frac{(1-\frac{2}{M}){\rm tr}(\hat{\bf
D}(i)\hat{\bf S}^*(i)) + {\rm tr}(\hat{\bf D}(i)){\rm tr}(\hat{\bf
D}^*(i))}{(i+1-\frac{2}{M}){\rm tr}(\hat{\bf D}(i)\hat{\bf
S}^*(i))+(1-\frac{i}{M}){\rm tr}(\hat{\bf D}(i)){\rm tr}(\hat{\bf D}^*(i))}$ \\
${\hat{\bf a}_1}(i)=\frac{{\bf P}\hat{\bf d}(i)}{{\lVert{{\bf P}\hat{\bf d}(i)}\rVert}_2}$ \\
$\hat{\nu}_0(i)={{\rm tr}(\hat{\bf R}(i))}/M$ \\
$\hat{\bf F}_0(i)={\hat{\nu}_0}(i){\bf I}$ \\
$\tilde{\bf R}(i)={\hat{\rho}_0}(i){\hat{\bf F}_0}(i)+(1-{\hat{\rho}_0}(i))\hat{\bf R}(i)$ \\
${\hat{\rho}_0}(i+1)=\frac{(1-{\frac{2}{M}}){\rm tr}({\tilde{\bf R}(i)}{\hat{\bf R}(i)})+{\rm tr}^2(\tilde{\bf R}(i))}{(i+1-{\frac{2}{M}}){\rm tr}({\tilde{\bf R}(i)}{\hat{\bf R}(i)})+(1-{\frac{i}{M}}){\rm tr}^2(\tilde{\bf R}(i))}$ \\
$\hat{\sigma}^2_1(i)=\frac{|{{\hat{\bf a}_1}^H}(i){\bf x}(i)|^2-{{\hat{\bf a}_1}^H}(i){\hat{\bf a}_1}(i){\sigma}^2_n}{|{{\hat{\bf a}_1}^H}(i){\hat{\bf a}_1}(i)|^2}$ \\
$\tilde{\bf R}(i)=\tilde{\bf R}(i)+{{\lVert{\tilde{\bf R}(i)}\rVert}_2} {\bf I}$ \\
${\tilde{\bf R}_{i+n}}(i)=\tilde{\bf R}(i)-\hat{\sigma}^2_1(i){\hat{\bf a}_1}(i){{\hat{\bf a}_1}^H}(i)$ \\
${{\tilde{\bf R}}_{i+n}}(i)={{\tilde{\bf R}}_{i+n}}(i) {\frac{2{\sigma}^2_n}{{\lVert{\tilde{\bf R}_{i+n}(i)}\rVert}_2}}$ \\
$\hat{\bf w}(i)=\frac{{\tilde{\bf R}^{-1}_{i+n}}(i)\hat{\bf a}_1(i)}{{\hat{\bf a}_1}^H(i){\tilde{\bf R}^{-1}_{i+n}}(i)\hat{\bf a}_1(i)}$ \\
\hline
\end{tabular}
\end{center}
\end{table}\vspace{-1em}


\section{Simulations}

In our simulations, a uniform linear array (ULA) of $M=12$
omnidirectional sensors with a spacing of half wavelength is
considered. Three source signals include the desired signal which is
presumed to arrive at ${\theta}_1=10^\circ$ and two interferers
which are impinging on the antenna array from directions
${\theta}_2=50^\circ$ and ${\theta}_3=90^\circ$. The
signal-to-interference ratio (SIR) is fixed at 20dB. Only one
iteration is performed per snapshot and we employ $i=50$ snapshots
and 100 repetitions to obtain each point of the curves. The
beamformer computed with LOCSME is compared to existing beamformers
in terms of the output SINR. For the beamformers of \cite{r7},
\cite{r8}, \cite{r10}, \cite{r11} and the beamformer with LOCSME,
the angular sector is chosen as
$[{\theta}_1-5^\circ,{\theta}_1+5^\circ]$ and $p=8$ principal
eigenvectors are used. The number of eigenvectors of the subspace
projection matrix $p$ is selected manually with the help of
simulations. For the beamformers of \cite{r7}, \cite{r8}, \cite{r10}
and \cite{r11} which also require an optimization technique, the CVX
software is used. The SINR performance versus snapshots and SNR of
the algorithms is shown in Figs. 1 and 2 and the number of snapshots
is $50$ for the SINR versus SNR plots. The average
execution time of the algorithms in \cite{r7}, \cite{r8}, \cite{r10}
and \cite{r11} is around $0.3$ sec/snapshot, while LOCSME only
requires $0.021$ sec/snapshot.

\subsection{Mismatch due to Coherent Local Scattering}

In this case, the steering vector of the desired signal is affected by a local scattering effect and modeled as
\begin{equation}
{\bf a}={\bf p}+\sum\limits_{k=1}^4{e^{j{\varphi}_k}}{\bf b}({\theta}_k),
\end{equation}
where ${\bf p}$ corresponds to the direct path while ${\bf
b}({\theta}_k)(k=1,2,3,4)$ corresponds to the scattered paths. The
angles ${\theta}_k(k=1, 2, 3, 4)$ are randomly and independently
drawn in each simulation run from a uniform generator with mean
$10^\circ$ and standard deviation $2^\circ$. The angles
${\varphi}_k(k=1, 2, 3, 4)$ are independently and uniformly taken
from the interval $[0,2\pi]$ in each simulation run. Notice that
${\theta}_k$ and ${\varphi}_k$ change from trials while remaining
constant over snapshots \cite{r3}. Figs. 1 (a) and 2 (a) illustrate
the SINR performance versus snapshots and SNR under the coherent
scattering case. LOCSME outperforms the other algorithms and is
close to the optimum SINR.

\subsection{ Mismatch due to Incoherent Local Scattering}

In the incoherent local scattering case, the desired signal has a
time-varying signature and the steering vector is modeled by
\begin{equation}
{\bf a}(i)=s_0(i){\bf p}+\sum\limits_{k=1}^4{s_k(i)}{\bf b}({\theta}_k),
\end{equation}
where $s_k(i)(k=0, 1, 2, 3, 4)$ are i.i.d zero mean complex Gaussian
random variables independently drawn from a random generator. The
angles ${\theta}_k(k=0, 1, 2, 3, 4)$ are drawn independently in each
simulation run from a uniform generator with mean $10^\circ$ and
standard deviation $2^\circ$. This time, $s_k(i)$ changes both from
run to run and from snapshot to snapshot. Figs. 1 (b) and 2 (b)
depict the SINR performance versus snapshots and SNR. Compared to
the coherent scattering results, all the algorithms have a
performance degradation due to the effect of incoherent local
scattering. However, LOCSME is able to outperform the remaining
robust beamformers over a wide range of input SNR. The reason for
the improved performance of LOCSME is the combined use of accurate
estimates of the INC matrix and of the steering vector mismatch.

Further testing with a larger number of antenna array elements
indicates that the performance of all algorithms degrades (e.g.
LOCSME has around 2dB degradation when $M=60$). In addition,
inappropriate choice for the angular sector in which the desired
signal is assumed to be located will lead to obvious performance
degradation.

\vspace{-1.0em}

\begin{figure}[!htb]
\begin{center}
\def\epsfsize#1#2{0.875\columnwidth}
\epsfbox{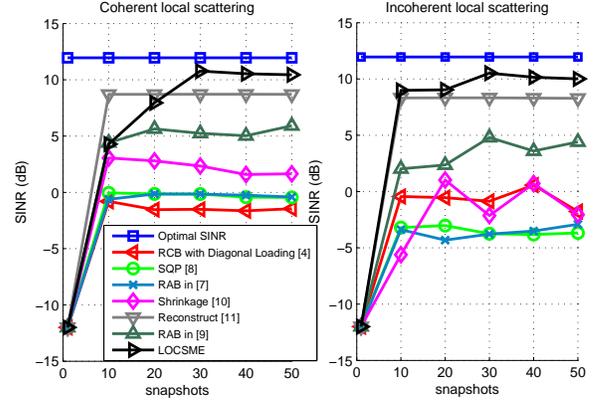} \vspace*{-1.2em} \caption{SINR versus snapshots.}
\label{1}
\end{center}
\end{figure}

\vspace{-1.0em}

\begin{figure}[!htb]
\begin{center}
\def\epsfsize#1#2{0.875\columnwidth}
\epsfbox{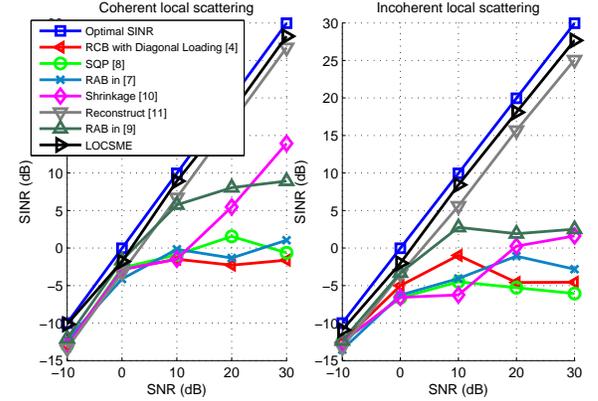} \vspace*{-1.2em} \caption{SINR versus SNR.}
\label{2}
\end{center}
\end{figure}

\vspace{1.4em}

\section{Conclusion}

We have proposed LOCSME that only requires prior knowledge of the
angular sector of the desired signal and is less costly than
existing methods. Simulation results have shown that LOCSME
outperforms prior art in both coherent local scattering and
incoherent local scattering cases.


\begin{appendix}
\textit{Derivation of $\hat{\rho}(i)$}: Equation (12) can be
rewritten in an alternative way in matrix version as $\hat{\bf
D}(i)=\hat{\rho}(i)\hat{\bf F}(i)+(1-\hat{\rho}(i))\hat{\bf S}(i)$.
By using (10), then the shrinkage intensity $\hat{\rho}(i)$ can be
computed from the following optimization problem \vspace{-0.5em}
\begin{equation}
\begin{aligned}
& \underset{\hat{\rho}(i), \hat{\nu}(i)} {\text{min}}
&& E[{\lVert{\hat{\bf D}(i)-\hat{\bf F}(i-1)}\rVert}^2] \\
& \text{subject to}
&& \hat{\bf D}(i)=\hat{\rho}(i)\hat{\nu}(i){\bf I}+(1-\hat{\rho}(i))\hat{\bf S}(i).
\end{aligned}
\end{equation}
Since $E[\hat{\bf S}(i)]=E[\hat{\bf F}(i-1)]$, the objective
function in (30) can be rewritten as
$\hat{\rho}^2(i){\lVert{\hat{\bf F}(i-1)-\hat{\nu}(i){\bf
I}}\rVert}^2+(1-\hat{\rho}(i))^2E[{\lVert{\hat{\bf S}(i)-\hat{\bf
F}(i-1)}\rVert}^2]$ \cite{r13}. The optimal value of $\hat{\nu}(i)$
is obtained as the solution to a problem that does not depend on
$\hat{\rho}(i)$ as given by
$\underset{\hat{\nu}(i)}{\text{min}}{\lVert{\hat{\bf
F}(i-1)-\hat{\nu}(i){\bf I}}\rVert}^2$, which can be solved by
computing the partial derivative of the argument with respect to
$\hat{\nu}(i)$ and equating the terms to zero. By substituting the
optimal value of $\hat{\nu}(i)$ into (30), computing the partial
derivative of the argument with respect to $\hat{\rho}(i)$, equating
the terms to zero and solving for $\hat{\rho}(i)$, we obtain
\vspace{-0.5em}
\begin{equation}
\hat{\rho}(i)=\frac{E[{\lVert{\hat{\bf S}(i)-\hat{\bf F}(i-1)}\rVert}^2]}{{\lVert{\hat{\bf F}(i-1)-\mu(i){\bf I}}\rVert}^2+E[{\lVert{\hat{\bf S}(i)-\hat{\bf F}(i-1)}\rVert}^2]}.
\end{equation}
By further Gaussian assumptions as in \cite{r12}, replacing
$\hat{\bf F}(i-1)$ by its estimate $\hat{\bf D}(i)$ and the data
sample number $n$ by the snapshot index $i$, equation (13) can be
obtained.
\end{appendix}

\end{document}